\documentstyle[aps]{revtex}
\begin{document}
\title{HIGH-FREQUENCY GRAVITATIONAL WAVES FROM \\ HAIRY BLACK HOLES}
\author{R.J. SLAGTER}
\address{Institute for Theoretical Physics, University of
Amsterdam, \\ Valckenierstraat 65, 1018XE Amsterdam, The Netherlands}
\maketitle
\abstract{
We investigated the stability of the spherically symmetric non-abelian
(Bartnik-McKinnon) black hole solution of the SU(2) Einstein-Yang-Mills
system  using the multiple-scale analysis.
It is found, in contrast with the vacuum situation, that a spherically
symmetric oscillatory perturbation to second order cannot be constructed.
The singular behavior of the gravitational waves is probably induced by the
coupling of the gravitational waves to the Yang-Mills waves.}
\section{Introduction}
The discovery of the non-abelian particle-like and non-Reissner-Nordstr\"om
(hairy) black hole solutions
of the SU(2) Einstein-Yang-Mills (EYM) theory,  first studied by Bartnik
and McKinnon (BMK)~\cite{bart}, initiated a throughout instability
investigation by several authors~\cite{strau1,lav,strau2,zhou}.
It turns out that under spherically symmetric perturbations the BM solution
is unstable. The nth BMK solution has in fact 2n unstable modes,
comparable with the flat spacetime "sphaleron" solution. In the static case,
it was recently conjectured~\cite{mav} that there are hairy black holes
in the SU(N) EYM-theories with topological instabilities.
In order to analyze the stability of the solutions, one usually linearizes
the field equations, or one expands the field variables in a
physically unclear small parameter~\cite{strau1,strau2}.
We conjecture that the conventional linear analysis is in fact inadequate
to be applied to the situation where a singularity is formed.
One better can apply the so-called multiple-scale or two-timing method,
developed by Choquet-Bruhat~\cite{choq1,choq2}. This method
is particularly useful for constructing uniformly valid approximations to
solutions of perturbation problems~\cite{slag1}. One of the advantages is that one
can keep track of the different orders of approximations. The method
originates from the Wentzel-Kramers-Brillouin(WKB)-method in order to construct
approximate solutions of the non-linear Schr\"odinger equation.
It can also be applied when there arise secular terms, as in the
Mathieu equation $ \frac{d^2y}{dt^2}+(a+2\epsilon \cos t)y=0.$
As a nice application of the multiple-scale-method, one finds the boundaries
between the regions in the $(a,\epsilon )$ plane for which all solutions are
stable and the regions in which there are unstable solutions.
Quite recently it was found that the method can be successful
for gravity with Gauss-Bonnet terms.
There is another application, namely the threshold of black hole
formation of the choptuon~\cite{chop1,chop2,garf,bizon}. Numerically it is found that there
is a critical parameter whose value separates solutions containing black holes
from those which do not. For a critical value one observes self-similarity:
it produces itself (echoes) on progressively finer scales (Choptuik scaling).
It is evident that the collapsing ball of field energy will produce
gravitational waves, which will be coupled to the  YM-field perturbations.
Due to the accumulation of echoes, the curvature diverges. The multiple-scale
method is suitable for handling this critical situation.

Let us consider a manifold ${\bf {\cal M}}$ with two different scales, i.e., a mapping
from ${\bf {\cal M}}\times {\cal R}$ into the space of metrics on
${\large{\cal M}}$:
\begin{equation}
\def\x{{\bf x}}
(\x,\xi)\Longrightarrow g(\x,\xi),\qquad \x\in{\cal M}, \xi \in {\cal R}.
\end{equation}
We set $\xi\equiv \omega \Pi({\bf x})$, with $\Pi$ a phase function on ${\cal M}$
of dimension of $length$ and $\omega$ a large parameter of dimension
$(length)^{-1}$.

The multiple-scale method assumes that the metric $g_{\mu\nu}$ and the
YM-potentials $A_\mu^a$ can be written as
\begin{equation}
g_{\mu\nu}=\bar g_{\mu\nu}+\frac{1}{\omega}h_{\mu\nu}(x^\sigma;\xi)+
\frac{1}{\omega^2}k_{\mu\nu}(x^\sigma;\xi)+... ,
\end{equation}
\begin{equation}
A_\mu^a=\bar A_\mu^a +\frac{1}{\omega}B_\mu^a(x^\sigma;\xi)+\frac{1}{\omega^2}
C_\mu^a(x^\sigma;\xi)+... ,
\end{equation}
where $\xi\equiv \omega\Pi(x^\sigma )$ and $\Pi$ a phase function.
The parameter $\omega$ measures the ratio of the fast scale to the slow one.
The rapid variation only occur in the direction of the vector
$l_\sigma \equiv \frac{\partial \Pi}{\partial x^\sigma}$. For a function
$\Psi(x^\sigma ;\xi )$ one has
\begin{equation}
\frac{\partial\Psi}{\partial x^\sigma}= \partial_\sigma \Psi
+\omega l_\sigma \dot \Psi,
\end{equation}
where $\partial_\sigma\Psi \equiv \frac{\partial \Psi}{\partial x^\sigma}
\vert_{\xi fixed}$ and $\dot\Psi \equiv \frac{\partial \Psi}{\partial \xi}\vert
_{x^\sigma fixed}$.
We consider here the case where the magnitude of the perturbation $h_{\mu\nu}$
of the metric with respect to the background $\bar g_{\mu\nu}$ is of order
$\omega^{-1}$ (for different possibilities, such as for example with the leading
term in the metric expansion of order $\omega^{-2}$, see
Taub~\cite{taub}).
Further, we will consider here the hypersurfaces $\Pi(x^\sigma )=cst $ as wave fronts
for the background metric $\bar g$, so (Eikonal equation)
\begin{equation}
l_\alpha l_\beta \bar g^{\alpha\beta}=0.
\end{equation}

Substituting the expansions of the field variables into the equations and
collecting terms of equal orders of $\omega$, one obtains propagation
equations for $\dot B_\mu^a, \ddot C_\mu^a, \dot h_{\mu\nu}$ and
$\ddot k_{\mu\nu}$ and 'back-reaction'
equations for $\bar g_{\mu\nu}$ and $\bar A_\mu^a$. It will be clear from the
propagation equation that there will be a coupling between the high-frequency
gravitational field and the high-frequency behavior of $A_\mu^a$ when the
singularity will be approached.
\section{ The EYM field equations in the Multiple-scale formulation}
Consider the Lagrangian of the SU(2) EYM system
\begin{equation}
S=\int d^4x\sqrt{-g}\Bigl[\frac{{\cal R}}{16\pi G}-
\frac{1}{4}{\cal F}_{\mu\nu}^a{\cal F}^{\mu\nu a}\Bigr],
\end{equation}
with the YM field strength
\begin{equation}
{\cal F}_{\mu\nu}^a =\partial_\mu A_\nu^a-\partial_\nu
A_\mu^a+g\epsilon^{abc}A_\mu^bA_\nu^c,
\end{equation}
g the gauge coupling constant, G Newton's constant, $A_\mu^a$ the gauge potential,
and ${\cal R}$ the curvature scalar. The field equations then become
\begin{equation}
G_{\mu\nu}=-8\pi G {\cal T}_{\mu\nu},
\end{equation}
\begin{equation}
{\cal D}_\mu {\cal F}^{\mu\nu a}=0,
\end{equation}
with ${\cal T}$ the energy-momentum tensor
\begin{equation}
{\cal T}_{\mu\nu}={\cal F}_{\mu\lambda}^a{\cal F}_\nu^{\lambda a}
-\frac{1}{4}g_{\mu\nu}{\cal F}_{\alpha\beta}^a {\cal F}^{\alpha\beta a},
\end{equation}
and ${\cal D}$ the gauge-covariant derivative, ${\cal D}_\alpha{\cal F}
_{\mu\nu}^a\equiv \nabla_\alpha{\cal F}_{\mu\nu}^a+g\epsilon^{abc}A_\alpha^b{\cal F}
_{\mu\nu}^c$.
Substituting the expansions into the YM-equation Eq. 9, we obtain for
the order $\omega$ equation
\begin{equation}
\bar g^{\mu\alpha}\Bigl[l_\mu l_\alpha \ddot B_\nu ^a -l_\nu \l_\alpha
\ddot B_ \mu ^a \Bigr] =0.
\end{equation}
For the order $\omega^0$-equation we obtain
\begin{eqnarray}
&&\bar g^{\mu\alpha} \Bigl[\bar\nabla_\alpha\bar {\cal F}_{\mu\nu}^a+
\Upsilon_{\alpha\mu}^\lambda \bar{\cal F}_{\nu\lambda}^a
- \Upsilon_{\alpha\nu}^\lambda \bar{\cal F}_{\mu\lambda}^a
+\dot B_\nu^a\Bigl(l_{\mu ,\alpha}-(\bar\Gamma_{\alpha\mu}^\lambda
+\Upsilon_{\alpha\mu}^\lambda )l_\lambda \Bigr)-\dot B_\mu^a\Bigl(l_{\nu ,\alpha} -
(\bar\Gamma_{\alpha\nu}^\lambda +\Upsilon_{\alpha\nu}^\lambda )l_\lambda\Bigr)\cr
&& +l_\mu \Bigl(\dot B_{\nu ,\alpha}-(\bar\Gamma_{\alpha\nu}^\lambda +
\Upsilon_{\alpha\nu}^\lambda )\dot B_\lambda^a\Bigr)
-l_\nu\Bigl(\dot B_{\mu ,\alpha}^a-(\bar\Gamma_{\alpha\mu}^\lambda
+\Upsilon_{\alpha\mu}^\lambda )\dot B_\lambda^a\Bigr)
+l_\alpha (\dot B_{\nu ,\mu}^a-\dot B_{\mu ,\nu}^a)
+l_\alpha (l_\mu \ddot C_\nu^a -l_\nu \ddot C_\mu^a)\cr
&&+g\epsilon^{abc}\Bigl(
l_\alpha (\bar A_\mu^b \dot B_\nu^c+\bar A_\nu^c \dot B_\mu^b)
+\bar A_\alpha^b (\bar {\cal F}_{\mu\nu}^c+l_\mu\dot B_\nu^c
-l_\nu\dot B_\mu^c )\Bigr)\Bigr] -h^{\mu\alpha} l_\alpha
(l_\mu\ddot B_\nu^a -l_\nu\ddot B_\mu^a)=0,
\end{eqnarray}
with $\Upsilon_{\mu\nu}^\lambda \equiv\frac{1}{2}\bar g^{\sigma\lambda}(l_\mu\dot h_{\nu\sigma}-
l_\nu\dot h_{\mu\sigma}-l_\sigma\dot h_{\mu\nu}).$

Substituting the expansions into the Einstein equations Eq. 8, we obtain
for the order $\omega$ equation

\begin{equation}
R_{\mu\nu}^{(-1)}=l_\nu\dot\Upsilon_{\mu\sigma}^\sigma -
l_\sigma\dot\Upsilon_{\mu\nu}^\sigma =0.
\end{equation}
For the $\omega^{0}$-equation we obtain
\begin{eqnarray}
\bar R_{\mu\nu}+R_{\mu\nu}^{(0)}=-8\pi G&&\Bigl\{\bar g^{\lambda\beta}(
\bar{\cal F}_{\mu\lambda}^a +l_\mu\dot B_\lambda ^a-l_\lambda\dot B_\mu^a)
(\bar{\cal F}_{\nu\beta}^a+l_\nu\dot B_\beta^a -l_\beta\dot B_\nu^a)
\cr &&+\frac{1}{4}\bar g_{\mu\nu}\bar g^{\sigma\alpha}\bar g^{\tau\beta}
(\bar{\cal F}_{\alpha\beta}^a+l_\alpha\dot B_\beta^a -l_\beta\dot B_\alpha^a)
(\bar{\cal F}_{\sigma\tau}^a+l_\sigma\dot B_\tau^a-l_\tau\dot B_\sigma^a)\Bigr\},
\end{eqnarray}
with $\bar R_{\mu\nu}$ the background Ricci tensor and $R_{\mu\nu}^{(0)}$ an
expression in $\ddot k, h\ddot h, \dot h^2$ and $\dot h$ (see Choquet-Bruhat
~\cite{choq1}).
We can simplify the equations considerable. If we use Eq. 5, then Eq. 11
becomes
\begin{equation}
l^\mu \ddot B_\mu^a =0,
\end{equation}
so for periodic $B_\mu^a$ we have
\begin{equation}
l^\mu B_\mu^a=0.
\end{equation}
Using Eq. 13 and Eq. 15  we obtain from Eq. 12
\begin{eqnarray}
\bar g^{\mu\alpha}\Bigl[\bar\nabla_\alpha\bar{\cal F}_{\mu\nu}^a-
\Upsilon_{\alpha\nu}^\lambda\bar{\cal F}_{\mu\lambda}^a +\dot B_\nu^a
\bar\nabla_\alpha l_\mu -\dot B_\mu^a\bar\nabla_\alpha l_\nu
+l_\mu\bar\nabla_\alpha\dot B_\nu^a-l_\nu\bar\nabla_\alpha\dot B_\mu^a
+l_\alpha(\dot B_{\nu,\mu}^a -\dot B_{\mu,\nu}^a )\Bigr] \cr
+l_\nu l_\alpha h^{\mu\alpha} \ddot B_\mu ^a
-l_\nu l^\mu \ddot C_\mu^a +g\epsilon^{abc}\bar g^{\mu\alpha}\Bigl[
\bar A_\alpha^b\bar{\cal F}_{\mu\nu}^c +2 l_\alpha \bar A_\mu^b\dot B_\nu^c
-l_\nu\bar A_\alpha^b\dot B_\mu^c\Bigr]=0,
\end{eqnarray}
with $\bar\nabla$ the covariant derivative with respect to the
background metric $\bar g_{\mu\nu}$.
Integrating this equation with respect to $\xi$ yields
\begin{equation}
\bar{\cal D}^\mu\bar{\cal F}_{\mu\nu}^a=0.
\end{equation}
Substituting back this equation into Eq. 12 we obtain the
propagation equation for the YM-field
\begin{eqnarray}
\bar\nabla^\mu (l_\mu\dot B_\nu^a-l_\nu \dot B_\mu^a)+l^\mu (\bar\nabla_\mu
\dot B_\nu^a -\bar\nabla_\nu\dot B_\mu^a)
-\bar g^{\mu\alpha}\Upsilon_{\alpha\nu}^\lambda\bar {\cal F}_{\mu\lambda}^a
+l^\alpha l_\nu h_\alpha^\lambda\ddot B_\lambda^a \cr
-l_\nu l^\mu \ddot C_\mu^a
+g\epsilon^{abc}\bar g^{\mu\alpha}(2l_\alpha\bar A_\mu^b\dot B_\nu^c-
l_\nu\bar A_\alpha^b\dot B_\mu^c )=0.
\end{eqnarray}
Multiplying this propagation equation Eq. 19 with $\dot B^{\nu a}$,
we obtain the 'conservation'-equation
\begin{equation}
\bar\nabla_\alpha (l^\alpha \dot B_\nu^a\dot B^{\nu a} )=l^\mu\dot h_\nu^\lambda
\dot B^{\nu a}\bar{\cal F}_{\mu\lambda}^a-2g\epsilon^{abc}l^\mu
\dot B_\nu^c\dot B^{\nu a}\bar A_\mu^b,
\end{equation}
so $\dot B_\nu^a\dot B^{\nu a}$ is not conserved, unless the right-hand side
is zero.
Multiplying Eq. 19  with $l^\nu$, we obtain
\begin{equation}
l^\nu (l^\lambda\dot h_\nu^\mu -l^\mu\dot h_\nu^\lambda )
\bar{\cal F}_{\mu\lambda}^a =0.
\end{equation}
For the choice $l^\nu h_{\nu\mu}=0 $ and hence from Eq. 13 $h=0$, the
equations simplify again.\\
Integrating Eq. 14 with respect to $\xi$, we obtain
\begin{equation}
\bar R_{\mu\nu}=-\frac{1}{4\tau} l_\mu l_\nu \int(\dot h_\rho^\sigma
\dot h_\sigma^\rho -\frac{1}{2}\dot h^2) d\xi -\frac{8\pi G}{\tau} l_\mu l_\nu
\int\dot B^{\alpha a}\dot B_\alpha ^a d\xi -8\pi G\bar{\cal T}_{\mu\nu},
\end{equation}
with $\tau$ the period of high-frequency components. In the right-hand side
we have the back-reaction term due to the high-frequency gravitational
perturbation, the high-frequency YM-perturbations and the background energy-
momentum term $\bar{\cal T}$.
Substitution back Eq. 22 into Eq. 14, we obtain the perturbation equations for
$h_{\mu\nu}$
\begin{eqnarray}
R_{\mu\nu}^{(0)}=&&l_\mu l_\nu\Bigl[
\frac{1}{4\tau}\int(\dot h_\rho^\sigma\dot h_\sigma^\rho-\frac{1}{2}\dot h^2) d\xi
+\frac{8\pi G}{\tau}\int\dot B^{\alpha a}\dot B_\alpha^a d\xi\Bigr]
-8\pi G\Bigl[(l_\mu l_\nu \dot B^{\lambda a}\dot B_\lambda^a \cr
&&+\bar{\cal F}_{
\mu\lambda}^a (l_\nu\dot B^{\lambda a}-l^\lambda\dot B_\nu^a )+
\bar{\cal F}_{\nu\lambda}^a (l_\mu\dot B^{\lambda a}-l^\lambda\dot B_\mu^a )
+\frac{1}{2}\bar g_{\mu\nu}\bar F_{\alpha\beta}(l^\alpha\dot B^{\beta a}-
l^\beta\dot B^{\alpha a})\Bigr]
\end{eqnarray}

\section{Application to the spherical symmetric spacetime}
Let us consider the spherical symmetric spacetime
\begin{equation}
ds^2=-m(t,r)N(t,r)^2dt^2+\frac{1}{m(t,r)}dr^2+r^2(d\theta ^2+\sin^2\theta d\varphi^2),
\end{equation}
We have for the tetrad component: $l_\mu =(-N,\frac{1}{m},0,0)$ and
$l^\mu =(\frac{1}{mN},1,0,0)$.
We parameterize the SU(2) YM field in the 'polar' gauge as~\cite{strau1}
\begin{equation}
A=A_\mu dx^\mu=A_\mu^a\tau_a dx^\mu
={\cal A}_1\tau_3 dt++{\cal A}_2 \tau_3dr +({\cal W}_1\tau_1+{\cal W}_2\tau _2) d\theta +(\cot\theta\tau_3+
{\cal W}_1\tau_2-{\cal W}_2\tau_1)\sin\theta d\varphi,
\end{equation}
with ${\bf \tau}_a$ the spherical projections of the SU(2) generators.
In components
\begin{eqnarray}
&&A_t=\Bigl[0,0,\bar a+\frac{1}{\omega}a_1+\frac{1}{\omega^2}a_2+...\Bigr],\cr
&&A_r=\Bigl[0,0,\bar b+\frac{1}{\omega}b_1+\frac{1}{\omega^2}b_2+...,\Bigr]\cr
&&A_\theta =\Bigl[-(\bar w+\frac{1}{\omega}w_1+\frac{1}{\omega^2}w_2+...),
-(\bar v(u)+\frac{1}{\omega}v_1+\frac{1}{\omega^2}v_2 +...),0\Bigr],\cr
&&A_\varphi =\Bigl[\bar v+\frac{1}{\omega}v_1+\frac{1}{\omega}v_2 +..,
(-(\bar w+\frac{1}{\omega}w_1+\frac{1}{\omega^2}w_2+...)\sin\theta,-\cos\theta\Bigr],
\end{eqnarray}
where the perturbations depend on $t,r,\theta,\varphi$ and $\xi$. From Eq. 16 we
obtain the condition $b_1=-\frac{a_1}{mN}.$ From Eq. 13 we obtain some restrictions on
$ h_{\mu\nu}$: $h_{01}=-\frac{h_{00}}{2mN}-\frac{1}{2}mNh_{11}, h_{12}=
-\frac{h_{02}}{mN}$ and $h_{13}=-\frac{h_{03}}{mN}$.
We can now evaluate the propagation equations and back-reaction equations. From the
YM propagation equation Eq. 19 we obtain the set ($g=-1$)
\begin{equation}
\bar w^2+\bar v^2 =0, \qquad \bar b=\frac{-\bar a}{mN},\qquad \partial_tm
-m^2\partial_rN=0,
\end{equation}
\begin{equation}
\partial_t\dot w_1 +mN\partial_r \dot w_1 =0,
\end{equation}
\begin{eqnarray}
\partial_t\dot a_1 +mN\partial_r\dot a_1 -\dot a_1 (\frac{\partial_t m}{m}
+\frac{\partial_tN}{N})+N\ddot a_2+mN^2\ddot b_2\cr
+(\partial_t \bar a+mN\partial_r\bar a-\frac{\bar a\partial_tm}{m}
-\frac{\bar a\partial_tN}{N})(\frac{(\dot h_{00}-m^2N^2\dot h_{11})}{2mN^2}=0
\end{eqnarray}

The third equation in Eq. 27 is consistent with the result that the
divergence of the null-congruence of $l^\mu$ must be equals $\frac{2}{r}$, and
the fact that $\partial_{rt}\Pi =\partial_{tr}\Pi$.
The back-reaction equation Eq. 18 for the background YM field yields (from now on we set
$\bar v =\bar w =0$)
\begin{equation}
\partial_t^2 \bar b -\partial_t\bar b\frac{\partial_t N}{N}+\partial_r \bar a
\frac{\partial_t N}{N}-\partial_{tr} \bar a =0,
\end{equation}
For the propagation equations Eq. 23 of the high-frequency gravitational perturbations $h$
and $k$, we obtain
\begin{equation}
\partial_t\dot h_{22}+mN\partial_r\dot h_{22}-\frac{mN\dot h_{22}}{r}=0,
\end{equation}
\begin{equation}
\partial_t\dot h_{23}+mN\partial_r\dot h_{23}-\frac{mN\dot h_{23}}{r}=0,
\end{equation}
\begin{eqnarray}
\ddot k_{13}+\frac{1}{mN}\ddot k_{03} =-\frac{\partial_t\dot h_{03}}{mN^2}
+\frac{\partial_r\dot h_{03}}{N}-\Bigl(\frac{\partial_tm}{m^2N^2}+
\frac{\partial_tN}{mN^3}+\frac{2}{rN}\Bigr)\dot h_{03}
-\frac{\partial_\varphi\dot h_{00}}{2mN^2}+\frac{m\partial_\varphi\dot h_{11}}{2}\cr
+\frac{\partial_\varphi\dot h_{22}}{r^2}-\frac{\partial_\theta\dot h_{23}}{r^2}-
\frac{\dot h_{23}\cot\theta}{r^2}-\frac{16\pi G\cos\theta}{m^2N^3}(mN\partial_t\bar a
-\bar a N\partial_tm-\bar a m\partial_tN+m^2N^2\partial_r\bar a),
\end{eqnarray}
\begin{eqnarray}
\ddot k_{12}+\frac{1}{Mn}\ddot k_{02}=\Bigl(\frac{2}{rN}+\frac{\partial_tm}
{m^2N^2}+\frac{\partial_tN}{mN^3}\Bigr)\dot h_{02} -\frac{16\pi G\cot\theta}{r^2}
-\frac{1}{r^2}\partial_\theta\dot h_{22}-\frac{2\cot\theta}{r^2}\dot h_{22}\cr
-\frac{1}{r^2\sin^2\theta}\partial_\varphi\dot h_{23}
+\frac{m}{2}\partial_\theta\dot h_{11}-\frac{1}{N}\partial_r\dot h_{02}
-\frac{1}{mN^2}\partial_t\dot h_{02}-\frac{1}{2mN^2}\partial_\theta\dot h_{00},
\end{eqnarray}
\begin{eqnarray}
&&\ddot k_{22} +\frac{\ddot k_{33}}{\sin^2\theta}=2m^2r\dot h_{11}-32\pi G(
\dot v_1^2+\dot w_1^2)+\frac{2}{r^2}h_{22}\ddot h_{22}+\frac{2}{\sin^2\theta r^2}
h_{23}\ddot h_{23}-\frac{2}{N}\partial_\theta\dot h_{02}-\frac{2\cot\theta}{N}
\dot h_{02}\cr
&&+\frac{\dot h_{22}^2}{r^2}+\frac{\dot h_{23}^2}{r^2\sin^2\theta}
-\frac{2}{N\sin^2\theta}\partial_\varphi\dot h_{03}
+\frac{1}{r^2\tau}\int\Bigl(\dot h_{22}^2+\frac{\dot h_{23}^2}{\sin^2\theta}
\Bigr)d\xi +\frac{32\pi G}{\tau}\int(\dot w_1^2+\dot v_1^2)d\xi,
\end{eqnarray}
and
\begin{eqnarray}
\ddot k_{11}+\frac{2}{mN}\ddot k_{01}+\frac{1}{m^2N^2}\ddot k_{00}=
\Bigl(\frac{2}{m^2N^4}\partial_tN+\frac{3}{m^3N^3}\partial_tm+\frac{1}{m^2N^2}
\partial_rm\Bigr)\dot h_{00}\cr
+\bigl(\frac{1}{mN}\partial_tm+\partial_rm\Bigr)\dot h_{11}+\frac{1}{N}
\partial_t\dot h_{11}+m\partial_r\dot h_{11}-\frac{1}{m^2N^3}\partial_t\dot h_{00}
-\frac{1}{mN^2}\partial_r\dot h_{00}.
\end{eqnarray}
From the back-reaction equations Eq. 22 we obtain
\begin{equation}
\partial_tm=\frac{N}{2\tau r^3}\int\Bigl(\frac{\dot h_{22}^2\sin^2\theta+
\dot h_{23}^2}{\sin^2\theta}\Bigr)d\xi+\frac{16\pi G N}{r}\int\Bigl(\frac{
\dot w_1^2\sin^2\theta+\dot v_1^2\sin^2\theta+\frac{1}{2}
\cos^2\theta}{\sin^2\theta}\Bigr)d\xi ,
\end{equation}
which represents the effect on the decrease of the metric ("mass") component
$M$ due to the high-frequency perturbations, where M is the well known mass
parameter in the Schwarzschild parameterization $m\equiv(1-\frac{2M}{r})$.

\section{Conclusions}
Let us try to find a bounded solution to second order.
Following the vacuum situation of Choquet-Bruhat~\cite{choq1}, we consider
the simplified situation, where we take $h_{02}=h_{03}=\ddot k_{02}=
\ddot k_{03}=0$ and $h_{\mu\nu}$ independent of $\varphi$. Further,
\begin{eqnarray}
&& \ddot k_{33}=-\sin^2\theta\ddot k_{22} \cr
&& h_{22}=r{\cal H}_{22}(r,t)\cos\xi\sin\theta \cr
&& h_{23}=r{\cal H}_{23}(r,t)\sin\xi\sin\theta\cos\theta.
\end{eqnarray}
From Eq. 31, Eq. 32 and Eq. 35  we could find a bounded (not spherically symmetric)
solution for the other components of $\dot h_{\mu\nu}$.
However, from Eq. 34 we conclude that $\ddot k_{12}$ is unbounded at the poles.

In a future work we will investigate the axially symmetric case, as initiated
before~\cite{slag2}

\end{document}